\documentclass{amsart}
\usepackage{amssymb}
\title{Integrable Systems and Topology of Isospectral Manifolds}
\author{Alexei V. Penskoi}
\address{Independent University of Moscow,
Bolshoy Vlasyevskiy per. 11,
119002 Moscow Russia \& 
Bauman Moscow State Technical University, Moscow, Russia.}
\email{penskoi@mccme.ru}
\date{}
\newtheorem{Proposition}{Proposition}
\DeclareMathOperator{\diag}{diag}
\DeclareMathOperator{\tr}{tr}
\DeclareMathOperator{\sgn}{sgn}
\begin{document}
\maketitle

\section*{Introduction}

Let us consider a symplectic manifold $(X^{2n}\!,\omega)$ and an integrable
system with a Hamiltonian $H$ and involutive integrals
$F_1=H, F_2,\dots,F_n.$ Let $X_F\subset X$ be a subset defined
by equations $F_1=k_1,\dots,F_n=k_n,$ where $k_i$ are constants.
The subset $X_F$ is a submanifold for generic $k_1,\dots,k_n.$
This submanifold is called a level surface of integrals.
The well-known Liouville-Arnold theorem~\cite{Arnold}
says that if $X_F$ is compact and connected, then it is a torus.
This makes investigating the topology of $X_F$ trivial. However, it turns out that in
some important examples of integrable systems such
a submanifold is compact but its topology is quite complicated.
This is due to the fact that in these examples
$X$ has points where
either $H$ is singular or $\omega$ is singular or degenerate.
In such situations the Liouville-Arnold theorem does not apply.
However, sometimes it is possible to define the corresponding flow on the whole
submanifold $X_F$ and use it to investigate the  topology of $X_F.$

For example, it is possible if 
an integrable system has a Lax representation $\frac{dL}{dt}=[L,A].$
It is clear that the corresponding flow is defined on the whole submanifold
$X_F.$ In this case this submanifold is an isospectral manifold
of Lax matrices $L$ defined by the equations $\text{tr}L^i=k_i.$
This leads us to a natural idea to investigate the topology of
isospectral manifolds of Lax matrices $L$ using the integrable
flows $\frac{dL}{dt}=[L,A].$ It turns out that in some examples 
the topology of these isospectral manifolds is interesting 
enough to study it.

\section{Topology of the isospectral variety of Jacobi matrices}

The first example of this kind was described in 1984 by Tomei in
his paper~\cite{T},
where the topology of the isospectral variety $J_n$ of Jacobi 
$n\times n$-matrices (i.e. real three-diagonal symmetric matrices) 
$$
\left(%
\begin{array}{ccccc}
a_1 & b_1 & 0 & \dots & 0 \\
b_1 & a_2 & b_2& \ddots & \vdots \\
0 & b_2 & \ddots & \ddots  & 0 \\
\vdots & \ddots& \ddots  & a_{n-1} & b_{n-1} \\
0 & \dots & 0 & b_{n-1}  & a_n \\
\end{array}\right)
$$
is investigated. 
This is a level surface of integrals
of the (open) Toda lattice
$$
\dot{a}_i=2(b^2_i-b^2_{i-1}),\quad i=1,\dots,n,
\quad\dot{b}_i=b_i(a_{i+1}-a_i),\quad i=1,\dots, n-1,\quad b_0=b_n=0.
$$
Using the Toda flow Tomei computed
the Euler characteristic of this variety
\begin{equation}\label{tomei}
\chi(J_n)=(-1)^{n+1}\frac{B_{n+1}2^{n+1}(2^{n+1}-1)}{n+1},
\end{equation}
where $B_k$ denotes the $k$-th Bernoulli number defined as
in~\cite{K}. 
Tomei also described in~\cite{T}
a CW-complex structure for the quotient of this variety
by the action of a discrete symmetry group.

Two years later Fried discovered~\cite{F} that the stable and unstable
stratifications defined by the Toda flow are cell complexes. 
Using this fact he found the Betti numbers of
this isospectral variety.

Let $\pi\in S_n$ be a permutation.
Recall that an interval
$[i,i+1]$ such that $\pi(i)<\pi(i+1)$ is called an ascent.
The number of permutations of $n$ elements with $k$
ascents is called the Euler number and denoted by $\genfrac{<}{>}{0pt}{}{n}{k},$
see~\cite{K}. This is an important combinatorial object.

Fried proved~\cite{F} that the Betti number $b_k$ is equal to
$\genfrac{<}{>}{0pt}{}{n}{k}.$
He also described the multiplication
in the cohomology ring.

This beautiful combination of a well-known integrable system, 
a non-trivial topology and combinatorics makes investigating
this isospectral variety interesting.

\section{Topology of the isospectral variety of zero-diagonal Jacobi matrices}

In a recent short communication~\cite{P1} new results concerning
the topology of the isospectral variety of zero-diagonal
Jacobi matrices were announced. Let us expose these results in
details. 

Consider the variety $M_k$ of all $k\times k$-matrices
of the form
\begin{equation}\label{L}
L=\left(%
\begin{array}{ccccc}
0 & c_1 & 0 & \dots & 0 \\
c_1 & 0 & c_2& \ddots & \vdots \\
0 & c_2 & \ddots & \ddots  & 0 \\
\vdots & \ddots& \ddots  & 0 & c_{k-1} \\
0 & \dots & 0 & c_{k-1}  & 0 \\
\end{array}\right)
\end{equation}
with fixed spectrum. 

\begin{Proposition}\label{spectrum}
a) The eigenvalues of $L$
are of the form $0$ (if $k$ is odd), 
$\pm\lambda_1,\pm\lambda_2,\dots$
b) If all eigenvalues are distinct
then $M_k$ is a compact smooth manifold and its topology
does not depend on the eigenvalues.
\end{Proposition}

\noindent{\bf Proof.} Multiplying all even rows and columns 
of $L-\lambda I$ by $-1,$ we obtain $-L-\lambda I.$ 
Hence $\det(L-\lambda I)=\det(-L-\lambda I);$
this implies a).
The proof of b) is analogous to Tomei's
proof of the analogous statement for the isospectral variety
of Jacobi matrices~\cite{T}. $\Box$

The manifold $M_k$ is a level surface of integrals
of the (open) Volterra system
$$
\dot{c}_i=\frac{1}{2}c_i(c^2_{i+1}-c^2_{i-1}),\quad c_0=c_{k}=0.
$$
Usually this system is written in terms of the variables  $u_i=c_i^2.$
It is well-known that the Volterra system can be written in the Lax form
$\dot{L}=[L,A(L)].$ It follows 
that the Volterra flow preserves the spectrum of $L,$
i.e. the Volterra flow is a flow on $M_k.$

The Volterra system is an integrable system. This was
proved in 1974 by Manakov~\cite{M} and independently
a year later by Kac and van Moerbeke~\cite{KvM}.
It is well known that the Volterra  system is bi-Hamiltonian,
see e.g.~\cite{FT}.

It turns out that from the point of
view of topology there are two different cases
depending on the parity of $k.$

Let $P_k(c_1,\dots,c_{k-1},\lambda)=\det(L-\lambda I)$
be a characteristic polynomial of a $k\times k$-matrix $L$
of the form~(\ref{L}). For example, $P_2(c_1,\lambda)=\lambda^2-c_1^2.$
Let us say that a set $J\subset\{1,\dots,k-1\}$ is totally disconnected
if $J$ does not contain two consecutive integers $i$ and $i+1.$ Let $I_0=1$ and
$I_l=\sum c^2_{i_1}\dots c^2_{i_l},$ $l\geqslant1,$
where summation is over all the $i_1<\dots<i_l,$
such that $\{i_1,\dots,i_l\}$ is totally disconnected.

\begin{Proposition} a) In the even $(k=2l)$ case
$$
P_{2l}(c_1,\dots,c_{2l-1},\lambda)=\sum_{i=0}^{l}(-1)^i\lambda^{2l-2i}I_i.
$$
b) In the odd $(k=2l+1)$ case
$$
P_{2l+1}(c_1,\dots,c_{2l},\lambda)=\sum_{i=0}^l(-1)^{i+1}\lambda^{2l+1-2i}I_i.
$$
c) The variety $M_k$ is defined by the equations $I_i=k_i,$ $i=0,\dots,l,$
where $k_i$ are constants.
\end{Proposition}

\noindent{\bf Proof.} Expanding $\det(L-\lambda I)$ by the last row and
the last column we obtain a recurrent formula
$$
P_N(c_1,\dots,c_{N-1},\lambda)=-\lambda P_{N-1}(c_1,\dots,c_{N-2},\lambda)%
-c^2_{N-1}P_{N-2}(c_1,\dots,c_{N-3},\lambda).
$$
Using this formula and identities $P_1(\lambda)=-\lambda$
and $P_2(c_1,\lambda)=\lambda^2-c_1^2$ we obtain
the statement of the proposition by induction. $\Box$

Let us remark that these formulae for $I_i$ is analogous to those
obtained in~\cite{VP} for the periodic Volterra system.

\begin{Proposition}\label{cases}
a) In the even $(k=2l)$ case $M_k$ has $2^l$ isomorphic
connected components. Moreover, each
component is isomorphic to an isospectral variety of Jacobi matrices.

b) In the odd $(k=2l+1)$ case $M_k$ is connected.

c) $M_k$ is compact.
\end{Proposition}

\noindent{\bf Proof.} If $k=2l$ then $I_l=c^2_1c^2_3\dots c^2_{2l-1}\ne0$
because all eigenvalues are distinct, see Proposition~\ref{spectrum}.
This implies that $M_k$ has at least $2^l$ components
defined by $l$ values $\sgn c_{2i-1},$ $i=1,\dots,l.$
The proof that each of these components is connected
is similar to Tomei's proof that
an isospectral variety
of Jacobi matrices is connected.

There exists a well-known map, see~\cite{V,D},
$$
a_i=\frac{1}{2}c_i(c^2_{i+1}-c^2_{i-1}),\quad b_i=-\frac{1}{2}c_{2i}c_{2i-1}
$$
transforming solutions of the Volterra system into solutions of
the Toda lattice.
If $k=2l$ then it is easy to check that this map is an isomorphism between
each connected component of $M_k$ and 
the corresponding isospectral variety of Jacobi matrices. This completes the proof of a).

The proof of b)  and c) is similar to Tomei's proof~\cite{T} that
the isospectral variety
of Jacobi matrices is connected and compact.
$\Box$

Proposition~\ref{cases} reduces studying the topology
of $M_k$ in the even case to the already known topology
of the isospectral variety of Jacobi matrices
studied by Tomei and Fried.
Let us consider only the odd case.

It is not clear now how to compute the homology groups.
The stable and unstable stratifications are not cell
complexes in this case, and it is not possible to apply a direct approach
similar to Fried's~\cite{F}. We should also say that
in 1992 Bloch, Brockett and Ratiu~\cite{BRR}
used the idea of the double bracket representation
to show that  the Toda flow is a gradient flow. 
It was shown in 1998~\cite{P} that the Volterra flow is also
a gradient flow. This leads us  to the natural idea
of finding the homology groups using the Morse complex. 
Unfortunately we cannot use the Morse complex for
computing the homology groups because the stable and unstable 
stratifications are not transversal to each other. The question
of computing the homology groups in the odd case remains open to the 
best of the author's knowledge.
In the present paper we compute the Euler characteristic 
of $M_k$ using the Volterra flow as it is
was announced in~\cite{P1}. The approach is similar to
Tomei's~\cite{T}, but the combinatorics of the equilibrium points
of the Volterra flow and their indexes are much more
complicated.

Let $K=\frac{1}{4}\diag(1,2,3,\dots)$ and $f(L)=\tr KL^2.$
It has been shown~\cite{P} that the (negative) gradient flow
of $f(L)$ with respect to some metric (explicitly
constructed in~\cite{P}) coincides with
the Volterra flow. Let us choose $\lambda_i$
(see Proposition~\ref{spectrum}) in such a way that
all $\lambda_i>0$ and $\lambda_i>\lambda_{i+1}.$

\begin{Proposition}\label{critical}
a) The critical points of $f(L)$ on $M_{2l+1}$ are in one-to-one
correspondence with triples $[j,s,\pi]$
consisting of a number $j$ with $0\leqslant j\leqslant l,$ an $l$-tuple
$s=(s_1,\dots,s_l)$ with each $s_i$ equal to $0$ or $1,$
and a permutation $\pi\in S_l.$ 

b) The index of a critical point
corresponding to $[j,s,\pi]$ is equal to the number of
indexes $0\leqslant i\leqslant l-1$ with $i\ne j,j-1$ and
$\pi(i)<\pi(i+1),$ plus $1$ if $j\ne l.$
\end{Proposition}

\noindent{\bf Proof.} The critical points are equilibrium points
of the Volterra flow. Using this fact and the explicit polynomial
equations $I_i=k_i$
of $M_k$ one can prove by induction that the critical points are
exactly those points with exactly $l$ of the $2l$ values $c_1,\dots,c_{2l}$ 
equal to zero and the additional property that $c_i\ne0$
implies $c_{i-1}=c_{i+1}=0.$
Looking at spectrum of the corresponding matrices one obtains the description
in the statement of the proposition. 
Let us write $c_1,\dots,c_{2l}$
as an $2l$-tuple $(c_1,\dots,c_{2l}).$ 
The triple $[0,s,\pi]$ corresponds to the critical point
$$(c_1,\dots,c_{2l})=%
(0,(-1)^{s_1}\lambda_{\pi(1)},0,(-1)^{s_2}\lambda_{\pi(2)},%
0,\dots,(-1)^{s_l}\lambda_{\pi(l)});$$
the triple $[l,s,\pi]$ corresponds to 
$$
((-1)^{s_1}\lambda_{\pi(1)},0,(-1)^{s_2}\lambda_{\pi(2)},%
0,\dots,(-1)^{s_l}\lambda_{\pi(l)},0);
$$
the triple $[j,s,\pi]$ with $j\ne0,l$ corresponds to
$$((-1)^{s_1}\lambda_{\pi(1)},0,(-1)^{s_2}\lambda_{\pi(2)},%
0\dots,(-1)^{s_j}\lambda_{\pi(j)},0,0,(-1)^{s_{j+1}}\lambda_{\pi(j+1)},%
0,\dots, (-1)^{s_l}\lambda_{\pi(l)}).$$ 
This proves a).
The formula b) for the index may be 
obtained by studying the Hessian
of $f(L).$ There are two ways to compute the
index, but we omit details because both ways
are quite long and technical. The first way
is to introduce good local coordinates and calculate the index
in a straightforward manner. The second way is
the following. Remark that $f(L)=\tr KL^2$
is defined for any $k\times k$-matrix $L.$
The idea is to consider $M_k$ as
a subvariety of an orbit $O$ of the group $GL(k)$ acting on
$k\times k$-matrices by conjugation. One should
start by studying the Hessian of the restriction $f|_O,$
and then pass to $f|_{M_k}.$
$\Box$

\begin{Proposition}
a) The Euler characteristic of $M_{2l+1}$ is equal to
\begin{equation}\label{pp}
\chi(M_{2l+1})=2^{2l+2}(2^{l+2}-1)\frac{B_{l+2}}{l+2},
\end{equation}
where $B_{l+2}$ is the corresponding Bernoulli number.

b) If we define $\chi(M_1)$ to be $0,$ then the exponential generating function
of numbers $\chi(M_{2l+1})$ is equal to $-\tanh^2(2z),$ i.e.
$$
-\tanh^2(2z)=\sum_{l\geqslant0}\chi(M_{2l+1})\frac{z^l}{l!}.
$$
\end{Proposition}

\noindent{\bf Proof.} 
Denote the number of ascents in a permutation $\pi$ by $p(\pi).$
Let $\psi(n)=\sum\limits_{m=0}^n(-1)^m\genfrac{<}{>}{0pt}{}{n}{k},$
then formula (7.56) in~\cite{K} implies that
\begin{equation}\label{tanh}
1+\tanh z=\sum\limits_{n\geqslant0}\psi(n)\frac{z^n}{n!}.
\end{equation}
Using Proposition~\ref{critical}, one can prove that the Euler
characteristic $\chi(M_{2l+1})$ is equal to
$$
\chi(M_{2l+1})=2^l\sum\limits_{j=1}^{l-1}\binom{l}{j}%
\sum_{\pi_1\in S_j,\pi_2\in S_{l-j}}(-1)^{p(\pi_1)+p(\pi_2)+1}.
$$
This huge expression can be rewritten in terms of $\psi(n)$ in the following way.
$$
\chi(M_{2l+1})=-2^l\sum\limits_{j=1}^{l-1}\binom{l}{j}%
\Bigl(\sum_{\pi_1\in S_j}(-1)^{p(\pi_1)}\Bigr)%
\Bigl(\sum_{\pi_2\in S_{l-j}}(-1)^{p(\pi_2)}\Bigr)=%
-2^l\sum_{j=1}^{l-1}\binom{l}{j}\psi(j)\psi(l-j).
$$ 
This identity and the identity~(\ref{tanh}) imply b).
Using the expansion of $\tanh z$ and the formula $\tanh'z=1-\tanh^2z,$
one obtains the formula~(\ref{pp}) from a).
$\Box$

\section{Conclusion}

We obtained formula~(\ref{pp}) for the Euler characteristic of the isospectral
variety $M_k$ of zero-diagonal real Jacobi $k\times k$-matrices for odd $k.$
This formula
shows that the topology of $M_k$ is quite complicated. For example,
one can see from this formula that the variety $M_5$
is a compact surface of genus 5. It could seem quite surprising
because this variety is defined by simple equations
$$
c_1^2+c_2^2+c_3^2+c_4^2=k_1,\quad c_1^2c_3^2+c_1^2c_4^2+c_2^2c_4^2=k_2
$$
in the space $\mathbb{R}^4$ of variables $c_1,\dots,c_4.$
It is very interesting to obtain the Betti numbers for $M_k.$
The answer is expected to include non-trivial combinatorics.

It is interesting to remark that the Euler characteristics of
isospectral varieties of Jacobi matrices~(\ref{tomei}) and zero-diagonal
Jacobi matrices~(\ref{pp}) are related by a simple formula
$$
\chi(J_{l+1})=(-1)^{l}2^l\chi(M_{2l+1}).
$$
It could be
interesting to explain this fact.

The author is indebted to A.~P.~Veselov for attracting his
attention to this problem, and to O.~Cornea and A.~Medvedovsky for 
fruitful discussions.

\end{document}